\documentclass[12pt]{iopart}
\usepackage[dvips]{graphicx}
\usepackage{amsfonts}
\usepackage{amssymb}
\usepackage{appendix}

\def\mn#1{\langle #1 \rangle}

\begin{document}

\title[Experimental amplification of an entangled photon: what if the detection loophole is ignored?]{Experimental amplification of an entangled photon: what if the detection loophole is ignored?}

\author{Enrico Pomarico, Bruno Sanguinetti, Pavel Sekatski, Hugo Zbinden, Nicolas Gisin}
\address{Group of Applied Physics, University of Geneva, 1211 Geneva,
Switzerland.}

\ead{enrico.pomarico@unige.ch}

\begin{abstract}
The experimental verification of quantum features, such as entanglement, at large scales is extremely challenging because of environment-induced decoherence. Indeed, measurement techniques for demonstrating the quantumness of multiparticle systems in the presence of losses are difficult to define and, if not sufficiently accurate, they provide wrong conclusions.
We present a Bell test where one photon of an entangled pair is amplified and then detected by threshold detectors, whose signals undergo postselection. The amplification is performed by a classical machine, which produces a fully separable micro-macro state.
However, by adopting such a technique, one can surprisingly observe a violation of the CHSH inequality. This is due to the fact that ignoring the detection loophole, opened by the postselection and the system losses, can lead to misinterpretations, such as claiming the micro-macro entanglement in a setup where evidently there is not.
By using threshold detectors and postselection, one can only infer the entanglement of the initial pair of photons, so micro-micro entanglement, as it is further confirmed by the violation of a non-separability criterion for bipartite systems. How to detect photonic micro-macro entanglement in the presence of losses with currently available technology remains an open question.
\end{abstract}

\newpage
\tableofcontents

\section{Introduction}
Typically, quantum effects, such as superposition and entanglement, can be observed exclusively in the microscopic regime. This limitation prevents quantum mechanics to easily access our everyday experience and to become intuitive.

Investigating quantum features at larger scales is a fundamentally relevant step, because it allows one to test the limits of quantum theory and to fully solve the conceptual difficulties of the quantum to classical transition, pointed out for example by the famous "Schr\"odinger's cat" paradox \cite{Schroedinger35}.

In recent years some quantum effects at mesoscopic scales have been observed. Schr\"odinger cat states
have been generated by making Rydberg atoms interact with a coherent field trapped in a high Q cavity \cite{Brune96},
by superposing localized wave packets of a single trapped $^{9}$Be$^{+}$ ion with distinct internal electronic states
\cite{Monroe96,Myatt00} or macroscopically distinct current states in superconducting quantum interference devices (SQUID) \cite{Friedman00} or also by subtracting photons to squeezed states \cite{Ourjoumtsev06}. Interference patterns for massive molecules, such as spherical fullerenes C$^{60}$, have also been reported \cite{Arndt99}.

Recently, the entanglement between a photon and a macroscopic field containing $\approx 10^4$ photons has been tested for the first time \cite{DeMartini08}. Photonic micro-macro entanglement can be in principle created by amplifying one photon initially belonging to an entangled pair via a phase covariant cloner, a unitary transformation that can be implemented with an optical parametric amplifier. This remarkable experiment also stimulated the idea of performing Bell tests with human eyes as detectors \cite{Sekatski09}, in order to explore ways of bringing the entanglement close to our sensible experience.

Nonetheless, the experimental exploration of quantum mechanics in the macroscopic domain is extremely challenging because multiparticle systems are more susceptible to interactions with the environment and can easily and quickly lose their quantum coherence \cite{Zurek03}. Experimental losses (including finite detector efficiency) can be seen as an interaction between the system and the environment and make the task of revealing the entanglement of multiparticle systems highly nontrivial.

In \cite{DeMartini08} the non separability of the photonic micro-macro entangled state is claimed by violating an entanglement witness criterion for a bipartite system, where the macro state is conceived as a qubit, and by adopting a postselection of the detected events based on a threshold condition, called orthogonality filter.
However, in \cite{Sekatski10} we have shown that decoherence results in a large increase of the dimension of the Hilbert space in which the macroscopic state is defined, bringing it far from its theoretical model of a qubit.

Therefore, verifying the existence of micro-macro entanglement after the losses requires measurements able to explore this Hilbert space, guaranteeing an extremely efficient photon number resolution. Coarse-grained techniques, such as threshold detectors of finite efficiency and postselection, are not sufficiently accurate and can lead to wrong conclusions.

In this paper we illustrate experimentally some issues occurring when the entanglement between a photon and a macroscopic field is investigated \cite{Sekatski10}.
We describe an experiment where one photon of an entangled pair is amplified inside a black box and then detected by threshold detectors, whose signals undergo a postselection that is independent of the measurement bases.
At first, the entanglement in the system can be tested by measuring a CHSH inequality that,
within the frame of quantum mechanics, can be considered just as an entanglement witness.

In a standard CHSH test~\cite{Clauser69}, the choice of the measurement basis is made before the photon is amplified, usually by an avalanche photodiode. Interestingly, in our setup, the measurement basis can be chosen \emph{after} the optical amplification.
We show that, even if the final micro-macro state is fully separable, a violation of the Bell inequality is surprisingly possible and drives one to claim micro-macro entanglement despite the fact that we prove that it is clearly not present. This result highlights the relevance of the detection loophole, opened by system losses and postselection: indeed, if it is not taken into account, misinterpretations can be provided from similar experiments.
By violating a second test of non-separability for bipartite systems, similar to the one measured in~\cite{DeMartini08}, we confirm that with threshold detectors and postselection, one can only demonstrate the initial entanglement between the two photons, prior to the amplification.
The experiment described in this paper can be even curiously carried out by using a human observer as a detector, as we show at the end of the paper.

\section{A CHSH-Bell test with the amplification of an entangled photon}\label{par:experiment}
\subsection{The idea of the experiment}

\begin{figure}[t]
\begin{center}
\includegraphics[width=0.7\textwidth]{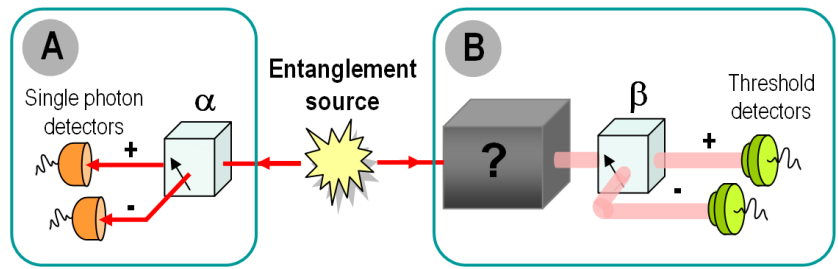}
\caption{Sketch of the idea of the experiment.}\label{fig:idea_experiment}
\end{center}
\end{figure}

Let us consider the following experiment.
Photon pairs entangled in polarization are distributed to two separated locations $A$ and $B$, as schematically sketched in figure \ref{fig:idea_experiment}.
The polarization of the photon on side $A$ is analyzed by a two-channel polarizer, measuring at an angle $\alpha$ with respect to the horizontal direction.
As in standard optical Bell tests, two single photon detectors are placed at the two outputs of the measuring device, labelled '+' and '-'.

The other photon of the pair is sent to a black box on side $B$.
For the moment we assume no knowledge of the internal functioning of the box, except the fact that for each incoming photon it sends a macroscopic pulse, thus performing an optical amplification.

Now, we would like to measure the CHSH inequality.
Actually, this is a rather unusual configuration for such a test. Indeed, in standard CHSH-Bell tests \cite{Clauser69}, each photon of the entangled pair
is measured in a specific measurement basis and then the result of the measurement is amplified
(electrically in a single photon detector) in order to be accessible by the experimenter. Here, the amplification is performed before the choice of the measurement basis.
By keeping the analogy with the CHSH test, the pulse coming from the box is analyzed by a two-channel polarizer as well, measuring at angle $\beta$ with respect to the horizontal direction. The outputs are labelled '+' and '-'.
The two orthogonally polarized pulses at the outputs of the polarizer have variable intensities. Each of the two orthogonal polarization components of this macroscopic pulse is incident on a threshold detector, which fires if the light is sufficiently intense.

A postselection technique is applied to the detected signals: if only one threshold detector fires, a conclusive result is produced. When both or none of them fire, the result is inconclusive and the event is rejected.
In order to perform a CHSH test, the correlations between the results on the two sides for some specific measurement bases have to be measured.

\subsection{The experimental implementation and the violation of the CHSH inequality} \label{par:experimental_implementation}

\begin{figure}[b]
\begin{center}
\includegraphics[width=0.9\textwidth]{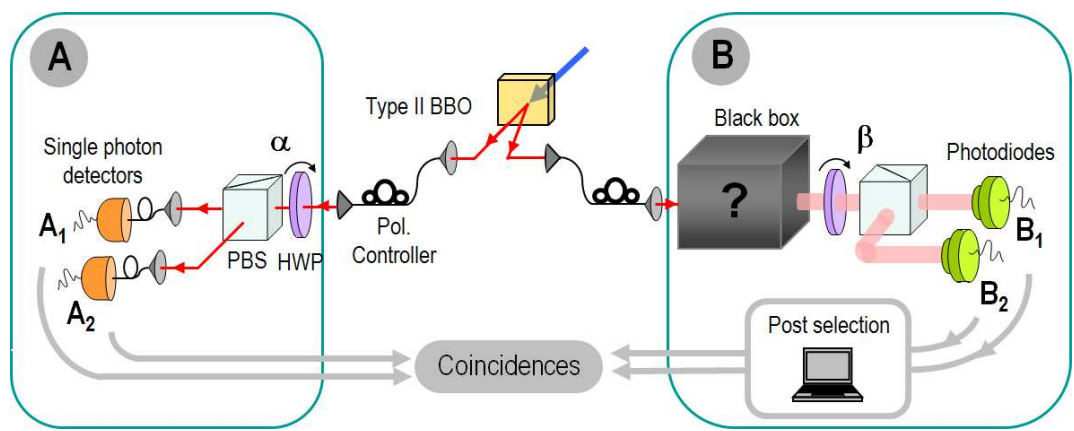}
\caption{Setup of the CHSH-Bell test with artificial threshold detectors made with photodiodes on side $B$.}\label{fig:scheme_exp}
\end{center}
\end{figure}

Photon pairs at 810\,nm entangled in polarization are produced by Spontaneous Parametric Down Conversion (SPDC) with a non linear BBO crystal cut for a type II phase matching and pumped by a diode laser at 405\,nm, as shown in figure \ref{fig:scheme_exp}. The pairs are generated in the singlet state $|\Psi^{-}\rangle=\frac{1}{\sqrt{2}}(|HV\rangle-|VH\rangle)$, where $|H(V)\rangle$ represents a single photon state of horizontal (vertical) polarization.

The photons cross interference filters which have a 10 nm bandwidth and are coupled into single mode fibres passing through polarization controllers.
The photon on side $A$ is analyzed in polarization with a system composed by a half wave plate (HWP) and a Polarizing Beam Splitter (PBS), which allows one to project into a linear polarization at an angle $\alpha$ with respect to the horizontal direction and to its orthogonal one at $\alpha^{\perp}$. The photons at the two outputs of the polarizer are detected by Avalanche PhotoDiode (APD) single photon detectors (Perkin Elmer), dubbed $A_1$ and $A_2$.

The photon on side $B$ is sent to the black box. The pulses emitted by the box are at 635\,nm and are similarly analyzed in polarization by a HWP and PBS, projecting into the polarizations at angles $\beta$ and $\beta^{\perp}$ with respect to the horizontal direction. Two linear photodiodes, dubbed $B_1$ and $B_2$, are adopted to detect the flashes at the two outputs of the PBS.
We record the analog output of these diodes and set, in software, a threshold voltage value on the detected signals, then the postselection technique described in the previous section is applied.
When only one of the signals overcomes the threshold, a conclusive result is produced.
The outcome is inconclusive and rejected when the signals are both above or both below the threshold value.

To violate the CHSH inequality, we measure the correlations of the results on the two sides using the following measurement bases.
On side $A$ we measure the polarizations at the angles ($\alpha_1=22.5^{\circ}$, $\alpha_1^{\perp}=112.5^{\circ}$) for the first basis $a_1$ or ($\alpha_2=67.5^{\circ}$, $\alpha_2^{\perp}=157.5^{\circ}$) for the second basis $a_2$. On side $B$ we measure at the angles ($\beta_1=0^{\circ}$, $\beta_1^{\perp}=90^{\circ}$) for the (H,V) basis ($b_1$) or ($\beta_2=45^{\circ}$, $\beta_2^{\perp}=135^{\circ}$) for the (+, -) one ($b_2$).
For each correlation term we take into account, in a total number of 5000 trials (each trial corresponding to the emission of a flash by the box), the coincidences between the single photon detectors on side $A$ and the conclusive answers provided by the threshold detectors on the side $B$.
The way in which coincidences are measured will be detailed in the following sections.
We need to estimate the Bell parameter
\begin{equation}
S=|E(a_1,b_1)-E(a_1,b_2)+E(a_2,b_1)+E(a_2,b_2)|\le 2,
\end{equation}
where $E(a_i,b_j)$ with $i,j=1,2$ is the correlation term in the basis $a_i$ and $b_j$ and $2$ represents the local bound.

By fixing a low threshold value such that we postselect 20\% of the detected events, a value of $2.45\pm 0.08$, which violates the local bound $2$, is measured for the Bell parameter.
Note that the 20\% ratio of detected events is independent of the choices of the measurement bases on $A$ and $B$ sides.
Now, we know from textbooks of quantum mechanics that a violation of a Bell inequality indicates the presence of entanglement.
But, what is exactly entangled with what? The pulse at the output of the black box with the single photon on the other side?

\section{The opening of the box}

The observed violation of the CHSH inequality could be explained if we assume having a Phase Covariant cloner \cite{Scarani05} in the black box. A Phase Covariant cloner is an optimal quantum cloning machine for qubits on the equatorial plane of the Poincar\'e sphere. Indeed, assuming a classical pump field, this cloner represents a unitary process which thus preserves the entanglement of the initial pair of photons \cite{Sekatski10}.

\begin{figure}[t]
\begin{center}
\includegraphics[width=1\textwidth]{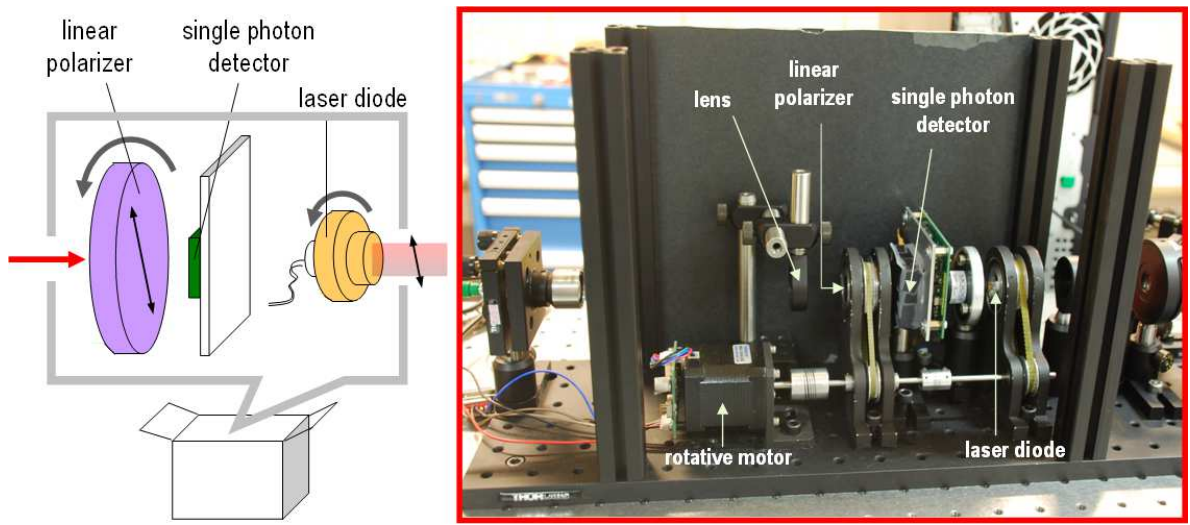}
\caption{Left: scheme of the Measure and Prepare cloner installed inside the black box. Right: picture of the internal structure of the Measure and Prepare cloner.}\label{fig:measure_and_prepare_cloner}
\end{center}
\end{figure}

However, when the box is opened, a much simpler machine is found.
As shown in figure \ref{fig:measure_and_prepare_cloner}, the incoming photons are focused onto a single photon detector by a converging lens placed behind a linear polarizer. The single photon detector (ID101) has 7\% of efficiency and low noise (0.7\,Hz). The polarizer is continuously rotated by an external motor, so the photons are measured in a random linear polarization. When the detector fires, a pulse is emitted by a diode laser, which is also rotated by the motor. The polarization of the pulse is aligned with the direction of the polarizer. As one photon of the pairs in the singlet state is unpolarized, a flash is sent at the output of the box with a random polarization and a constant probability corresponding to the efficiency of the detector.

This amplifying machine, which is a Measure and Prepare cloner, is absolutely classical: the state of the photon entering the black box is projected onto a random basis on a great circle of the Poincar\'e sphere and, if the detector clicks, the measured state is amplified.
It is very similar to the Phase Covariant cloner because the amplification is limited only to the great circle of the Poincar\'e sphere and both machines clone one qubit of the plane into a large number of qubits with a fidelity of $\frac{3}{4}$ \cite{Bae06}. Moreover, the Measure and Prepare cloner can be designed to emit photon states providing the same measurement statistics of the Phase Covariant cloner. Hence, it can be considered a sort of classical simulation of the quantum cloner.

However, the two devices, beyond their practical implementation, strongly differ in one fundamental point. The Phase Covariant cloner performs a unitary transformation on the initial quantum state and thus can in principle transfer the entanglement of two initial photons to the micro-macro regime.
On the contrary, the Measure and Prepare cloner installed in the black box performs a detection, therefore it breaks completely the entanglement between the two initially entangled photons.

It is clear now that a violation of a CHSH inequality by means of a threshold detection system and a postselection processing procedure, does not allow one to distinguish between a Phase Covariant and a Measure and Prepare cloner.
Therefore, the violation we reported in the previous paragraph does not reveal any kind of micro-macro entanglement.

\subsection{The importance of the detection loophole}

\begin{figure}[t]
\begin{center}
\includegraphics[width=0.7\textwidth]{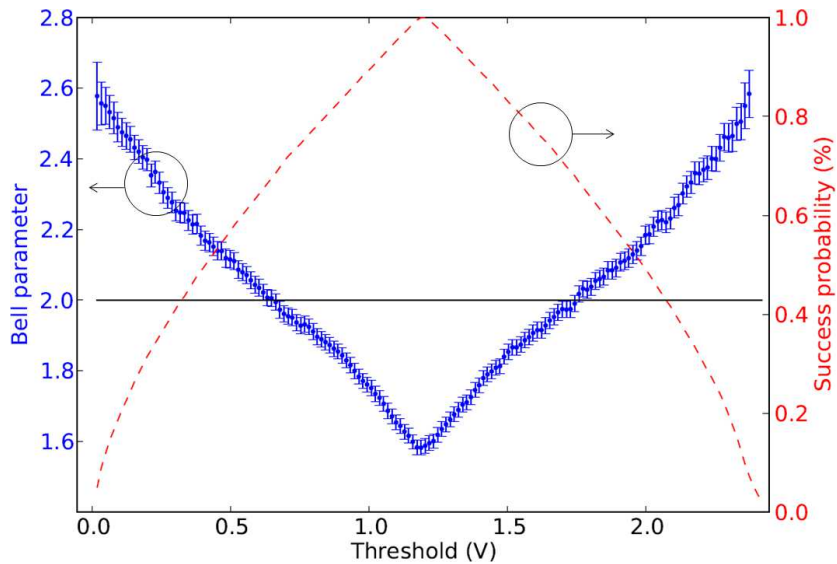}
\caption{Bell parameter (blue points) and success probability (red dashed line) as a function of the threshold value on the electrical signals provided by the two photodiodes.}\label{fig:S VS threshold}
\end{center}
\end{figure}

We repeat the Bell test by varying the threshold value on the signal recorded by the photodiodes on side $B$.
In figure \ref{fig:S VS threshold} the values of the Bell parameter are reported as a function of different thresholds.
The local bound at 2 for the CHSH inequality is represented by a horizontal line.

A violation of the local bound is evident only for small and large values of the threshold.
The two electrical signals are often above the threshold in the case of small thresholds, and below the threshold for large ones. In these cases several events are rejected and so a large amount of postselection is used. For small or large thresholds the error in the Bell parameter increases because of the smaller sample of data with which it has been evaluated. Moreover, notice that the values of the Bell parameter are affected also by the imperfect nature of the singlet state produced by our source and obtained after the fibre transmission of the photons.

We quantify the postselection in terms of the success probability $P_s$, or probability of a conclusive result, that can be calculated as the ratio between the overall number of conclusive events and the total number of trials.
In figure \ref{fig:S VS threshold} the values of the success probability as a function of the threshold values are also
reported and represented by the red dashed line.

The violation of the CHSH inequality is large for small values of the success probability, so for a strong postselection.
This effect is related to the fact that a conclusive result is obtained when the polarization measured in the Measure and Prepare cloner is parallel to the one measured on side $B$. Therefore, increasing the amount of postselection means being more restrictive on this condition and accepting only the cases in which the two polarizations are the more and more parallel. Notice that the postselection efficiency is, on average, the same for all polarization measurement bases on side $B$ and is independent on the basis set on side $A$.

In our test the detection loophole is opened by the postselection and the system losses. This example shows that if the detection loophole is ignored, one can easily make wrong conclusions. In our case, one could claim that the macroscopic state produced by the black box is entangled with the single photon, but it is clearly not the case because the Measure and Prepare cloner gives rise to a completely separable state.

The result highlighted in this paragraph can be applied to general micro-macro entanglement witness criteria. When losses are present, micro-macro entanglement deteriorates and, if it is still present, is extremely tough to detect.
Indeed, coarse-grained measurement techniques, such as threshold detectors and postselection, are not able to capture the complexity of the Hilbert space in which the macrostate lives and the possible entangled nature of the photonic micro-macro system.

\subsection{Micro-micro entanglement}

In the previous paragraphs we have proved that threshold detectors and postselection turn out to be inappropriate to detect micro-macro entanglement. But does the measured violation of the Bell inequality in the presence of postselection provide other useful information?
In the context of quantum mechanics, a Bell inequality can be considered an entanglement witness even in the presence of losses.
Indeed, entanglement is a concept defined within Hilbert space quantum mechanics. Thus, an entanglement witness assumes a specific Hilbert space and can rely on hypotheses accepted by quantum physics, like the fact that inefficient detectors or, in general, losses independent of the measurement settings, do not modify the conclusions of an entanglement test \cite{Sekatski10}.
Therefore, the Bell inequality in the presence of postselection can be used to detect entanglement of a system composed by two qubits, so in a Hilbert space of dimension 2$\otimes$2 and, in our case, its measured violation shows the entanglement of the initial pair of photons.

In order to further confirm the micro-micro entanglement, we can consider the optical amplification performed by the Measure and Prepare cloner as a part of the detection system and measure an entanglement witness similar to the one measured in \cite{DeMartini08}.

According to this criterion for a separable state the following condition yields
\begin{equation}\label{eq:entanglement_witness}
|V_{a,a_{\perp}}+V_{a',a'_{\perp}}|\leq 1,
\end{equation}
where $V_{a,a_{\perp}}$ and $V_{a',a'_{\perp}}$ are the correlation visibilities in two orthogonal bases in a great circle of the Poincar\'e sphere $(a,a_{\perp})$ ans $(a',a'_{\perp})$. In the Appendix A a derivation of this entanglement witness can be found. The correlation visibility in the basis $(a,a_{\perp})$ can be measured by fixing the basis $(a,a_{\perp})$ for both sides $A$ and $B$ and calculating the quantity
\begin{equation}\label{eq:visibility}
V_{a,a_{\perp}}=\frac{N_{A_1,B_1}+N_{A_2,B_2}-N_{A_1,B_2}-N_{A_2,B_1}}{N_{A_1,B_1}+N_{A_2,B_2}+N_{A_1,B_2}+N_{A_2,B_1}},
\end{equation}
where $N_{i,j}$ are the coincidences between the detector $i=A_1$ ($A_2$) measuring the state $a$ ($a_{\perp}$) and the conclusive results provided by the photodiodes $j=B_1$ ($B_2$) measuring similarly the states $a$ ($a_{\perp}$).
\begin{figure}[t]
\begin{center}
\includegraphics[width=0.6\textwidth]{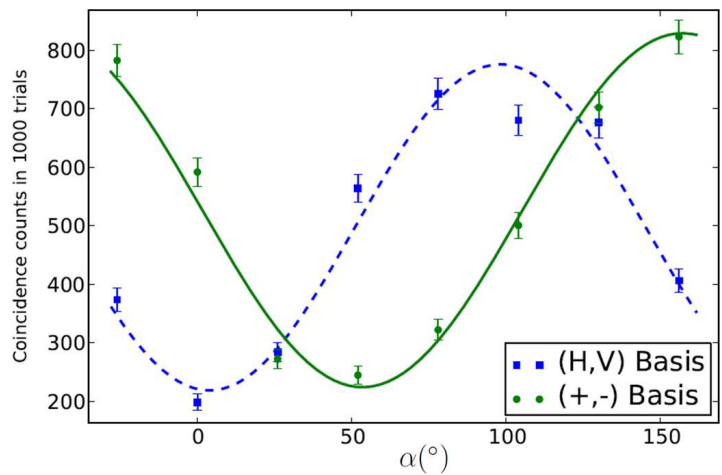}
\caption{Coincidence counts in 1000 trials between the detector $A_1$ and all results given by the photodiode $B_1$ for the bases ($H,V$) (square blue data with continuous fitting curve) and ($+,-$) (circle green data with dashed fitting curve). The fitting curves are used together with those of the coincidences between $A_1$ and $B_2$ for estimating the correlation visibilities.}\label{fig:basisHV+-}
\end{center}
\end{figure}
The Measure and Prepare cloner acts on the great circle of the Poincar\'e sphere corresponding to the linear polarizations, therefore two orthogonal bases like $(H,V)$ and $(+,-)$ can give non zero contributions to the total correlation visibility. In this case the bound in eq. (\ref{eq:entanglement_witness}) reads
\begin{equation}\label{eq:entanglement_witness2}
|V_{H,V}+V_{+,-}|\leq 1.
\end{equation}
In order to measure the visibility in the basis ($a,a_{\perp}$), we set this basis on side $B$ and change continuously the phase $\alpha$ on the photon on side $A$.
By using a threshold value of 1.2\,V, at which no postselection is performed, we observe interference fringes (figure \ref{fig:basisHV+-}) with visibilities of $0.536\pm 0.019$ for the $(H,V)$ basis and $0.602\pm 0.018$ for the $(+,-)$ basis. The difference between these two correlation visibilities is already present in the state of the two initially entangled photons, prior to the amplification.
The reported values give $|V_{H,V}+V_{+,-}|=1.138\pm0.026$, which violates the bound in eq. (\ref{eq:entanglement_witness}).

We observe that the entanglement witness is violated in a more consistent way if postselection is performed.
However, it is important to notice that, even if postselection of the detected signals is not necessary, the detected events are in any case only a part of the events generated by the entanglement source because of the inefficiency of the detectors, so the detection loophole is still open.

The confirmation of the existence of micro-micro entanglement puts in evidence another interesting result.
The Measure and Prepare cloner breaks the entanglement but, in some sense, not the information about it. Indeed, it is able to preserve the symmetry of the initial photon pair state \textit{after} the amplification.
If the measurement basis is chosen prior to the amplification, the information carried by the macroscopic state is simply the result of the measurement. While, when the amplification is performed before the measurement, the macroscopic state still carries the answers to all possible questions.

\section{Replacing artificial detectors with a human observer}

The CHSH-Bell test described in this paper can be in general carried out by adopting any kind of threshold detectors on side $B$.
Since the measure and prepare cloner optically amplifies the result of the measurement on a single photon to the classical regime,
visible with the naked eyes, threshold detection criteria based on the human visual system can be used.

\begin{figure}[t]
\begin{center}
\includegraphics[width=0.6\textwidth]{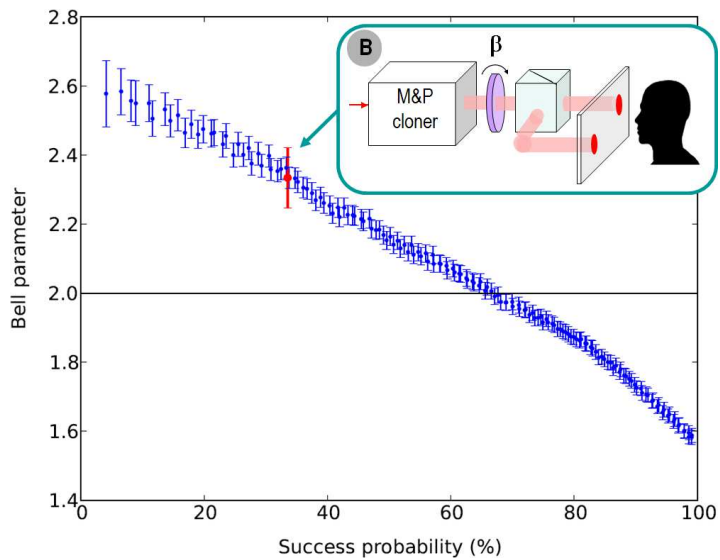}
\caption{Value of the Bell parameter as a function of the success probability for the measurement of the CHSH inequality with artificial threshold detectors (blue) and a human observer (red) on side $B$. Inset: configuration of side B with a human observer.}\label{fig:results_eye_and_scheme}
\end{center}
\end{figure}

Indeed, we perform a Bell test with one human observer on side $B$, replacing artificial detectors in a way that will be clarified below.
Similarly to the test described in \ref{par:experimental_implementation}, one photon of the entangled pair is amplified by the Measure and Prepare cloner and analyzed by a two-channel polarizer.
The two pulses at the output of the polarizer are sent to a piece of paper and the observer looks at the two spots produced on this (see inset of figure \ref{fig:results_eye_and_scheme}).
Since the polarization of the pulses is always changing, the intensity of the two spots varies. When the intensity of the two spots is quite different, the observer is asked to decide which spot is brighter by pushing a button corresponding to the left or to the right one. The labels '+' and '-' are associated to the two possibilities respectively.  However, he often cannot distinguish the two spots in intensity. In this case he is not able to give a conclusive answer, so he rejects the trial and waits for the following one. In other terms, he performs a postselection of the events in a way similar to what has been previously done by software on the electrical signals of the photodiodes.
The experiment is performed in a darkened room with a constant low level of luminosity and lasts approximately two hours. The pulses have a time duration of 200\,ms and a maximum peak power of approximately 1\,$\mu$W before the paper.

The data acquisition is carried out in the following way. Once the detector inside the cloner fires, the pulse is emitted, a sound alerts the observer to answer and the answers are registered by software together with the time-stamp of the detection inside the cloner.

We measure the CHSH inequality by setting the measurement bases as described in the paragraph \ref{par:experimental_implementation}.
For each of the four correlation terms to measure, we take into account the coincidences in 1000 trials between the single photon detectors $A_1$ and $A_2$ and the conclusive answers provided by the observer.
Notice that using two single photon detectors on side $A$ guarantees that the observer does not know which coincidence term of the correlation term he is measuring for each trial, avoiding any kind of arbitrariness in the results.
The measurement bases for the four correlation terms are also varied after every 250 trials in order to average fluctuations in time of the threshold criterion of the observer.

Table \ref{tab:coincidences} reports the values of the coincidences collected for the measurement of the CHSH inequality in this new configuration, providing the following values for the correlation terms:
\begin{table}[t]
 \centering
\begin{tabular}{l|cc|cc}
   & $\alpha_1$& $\alpha_1^{\perp}$  & $\alpha_2$ & $\alpha_2^{\perp}$ \\
   \hline

    $\beta_1$         & 15 $\pm$ 4  &  134 $\pm$ 12 & 112 $\pm$ 11  &  44 $\pm$ 7 \\
 $\beta_1^{\perp}$    & 144 $\pm$ 12  &  26 $\pm$ 5 & 46 $\pm$ 7  &  135 $\pm$ 12  \\
    \hline
   $\beta_2$          & 35 $\pm$ 6  &  132 $\pm$ 11 & 29$\pm$ 5  &  135 $\pm$ 12  \\
    $\beta_2^{\perp}$ & 118 $\pm$ 11  &  59 $\pm$ 8 & 150 $\pm$ 12  &  27 $\pm$ 5 \\
\end{tabular}
\caption{\small Coincidences values collected for the measurement of the CHSH inequality with one human observer on the side $B$ of the experiment.}
\label{tab:coincidences}
\end{table}
\begin{eqnarray}
E(a_1,b_1)&=& -0.743 \pm 0.038 \nonumber\\
E(a_1,b_2)&=& 0.466	\pm	0.048  \nonumber\\
E(a_2,b_1)&=& -0.453 \pm	0.048 \nonumber\\
E(a_2,b_2)&=& -0.672	\pm	0.040
\end{eqnarray}
These results give for the Bell parameter the value 2.334 $\pm$ 0.087, providing a violation of the local bound 2.
The average amount of postselection used in the experiment corresponds to a success probability of 33.5\%. The violation is in good agreement with the results given by the artificial threshold detectors, as shown in figure \ref{fig:results_eye_and_scheme}.\footnote{We also performed the same Bell test by using, instead of artificial detectors, two observers at each output of the PBS on side $B$ and by attenuating the pulses to approximately 1\,nW of peak power before the paper. This attenuation provided to the eyes an amount of light at the threshold vision level, roughly a few thousand photons scattered from the paper and impinging the observers's corneas. As for artificial threshold detectors, the events were conclusive only when one of the two observers saw the spot. We repeated these experiments with different pairs of observers, but the CHSH inequality was not systematically violated. In our opinion this was due to the fact that the visual detection thresholds of the two observers, related to individual states of concentration and light adaptation, vary in an independent way during a rather long experiment.}

It is important to stress again that in this kind of Bell test the human observer does not look at anything entangled but he is just witnessing the micro-micro entanglement present in the past in the system.
Adopting the human visual system for detecting a quantum phenomenon is fascinating and can be of interest for outdoor demonstrations or for highlighting the importance of the detection loophole in undergraduated laboratory courses.

\section{Discussion and conclusions}

In this paper we discuss some difficulties and limitations occurring in the experimental investigation of the entanglement between a photon and a macroscopic field.
We describe a CHSH-Bell test where one photon of an entangled pair is amplified by a Measure and Prepare cloner and detected by using threshold detectors and a postselection technique on the detected events. This is an unusual configuration for a Bell test, where the amplification is performed before the choice of the measurement bases.
A violation of the CHSH inequality is observed, although the final micro-macro state is completely separable.
This result emphasizes the importance of the detection loophole, which is opened by the postselection and the system losses, that is a serious problem that cannot be ignored.
This result is quite general. Detecting micro-macro entanglement in the presence of losses is extremely tough: coarse-grained measurement techniques, such as threshold detectors and postselection, are not sufficient. Extremely efficient photon number resolving detectors should be used together with appropriate entanglement witness criteria.

However, the experiment reported in this paper can show the entanglement of the two initial photons, so micro-micro entanglement. This has been shown by observing the violation of a non-separability criterion for bipartite systems, defined in \cite{Sekatski10}. The violation of the CHSH inequality can be even curiously obtained by replacing artificial threshold detectors with a human observer.
Adopting the human visual system for detecting a quantum phenomenon could be of interest for highlighting the importance of the detection loophole in undergraduated laboratory courses.

Detecting micro-macro entanglement in the presence of losses is an open issue that needs fundamental investigations, where new appropriate
criteria have to be formulated \cite{Spagnolo10,Spagnolo11} and important questions should find an answer: which amount of decoherence can be accepted for a genuine micro-macro entanglement test? Can one really detect photonic micro-macro entanglement with current available technology?

\addcontentsline{toc}{section}{Acknowledgments}
\section*{Acknowledgments}
We would like to acknowledge Fabio Sciarrino for useful discussions and Olivier Guinnard and Claudio Barreiro for the technical support.
Financial support for this project has been provided by the Swiss FNRS.

\begin{appendices}
\addcontentsline{toc}{section}{Appendix A: Derivation of the micro-micro entanglement witness}
\section*{Appendix A. Derivation of the micro-micro entanglement witness}
In \cite{Sekatski10} it has been used an entanglement witness criterion according to which for a separable state
$$|V_x +V_y+V_z|\leq1,$$
where $V_x$, $V_y$ and $V_z$ are the components of the correlation visibility of the state in three orthogonal bases in the Poincar\'e sphere.
It can be derived by taking into account the quantity $|\mn{\vec J_A \cdot \vec J_B}|$, where $J_A = a^\dag a - a_{\perp}^\dag a_{\perp}$ and $J_B = b^\dag b - b_{\perp}^\dag b_{\perp}$, that for a separable state $\rho_A\otimes\rho_b$ gives $|\mn{\vec J_A}_{\rho_A} \cdot\mn{ \vec J_B}_{\rho_B}|$.
An upper bound of this term can be found by observing that $\mn{\vec J_{A(B)}}$ is a (``classical'') vector and the scalar product between two vectors is smaller then the product of their lengths
\begin{equation}
|\mn{\vec J_A}_{\rho_A} \cdot \mn{\vec J_B}_{\rho_B}|\leq |\mn{\vec J_A}||\mn{\vec J_B}|\leq\mn{N_A}\mn{N_B}.
\end{equation}
This is basically the proof of the entanglement witness in \cite{Sekatski10}.

In the same fashion we can derive a witness for only two visibility components. We consider now the term $|\mn{ J_A^x  J_B^x + J_A^y  J_B^y}|$, where $x$ and $y$ stay for two arbitrary orthogonal measurement settings ($J_A^x = a_x^\dag a_x - a_{x\perp}^\dag a_{x\perp}$ ), which gives $|\mn{ J_A^x}_{\rho_A}\mn{  J_B^x}_{\rho_B} + \mn{J_A^y}_{\rho_A}  \mn{J_B^y}_{\rho_B}|$ for a separable state $\rho_A\otimes\rho_b$.
Now, this quantity can be seen as the scalar product $|P_{xy}\mn{\vec J_A}\cdot P_{xy}\mn{\vec J_B}|$, with $P_{xy}\mn{\vec J_{A(B)}}$ being the projection of $\mn{\vec J_{A(B)}}$ on the $x$-$y$ plane. These projected vectors have clearly a smaller length with respect to the initial vectors, thus we have
\begin{eqnarray}
|\mn{ J_A^x}_{\rho_A}\mn{  J_B^x}_{\rho_B} + \mn{J_A^y}_{\rho_A}  \mn{J_B^y}_{\rho_B}| \leq \nonumber\\|P_{xy}\mn{\vec J_A}||P_{xy}\mn{\vec J_B}|
\leq |\mn{\vec J_A}||\mn{\vec J_B}|\leq\mn{N_A}\mn{N_B}.
\end{eqnarray}
Using this entanglement witness we just have to measure two components of the correlation visibility and don't need any assumption on a third one.

\end{appendices}

\end{document}